# Ultrafast Chirality-dependent Dynamics from Helicity-resolved Transient Absorption Spectroscopy

Xiu Zhang[1,2,3,#], Lu Zhang[1,#], Junzhi Zhu[1,#], Tingxiao Qin[1], Haiyun Huang[1], Baixu Xiang[4], Haiyun Liu[1,*], and Qihua Xiong [4,5,6,*]

[1] Beijing Academy of Quantum Information Sciences, Beijing 100193, P.R. China;

[2] Beijing National Laboratory for Condensed Matter Physics, Institute of Physics, Chinese Academy of Sciences, Beijing 100190, P. R. China;

[3] University of Chinese Academy of Sciences, Beijing 100049, P. R. China;

[4] State Key Laboratory of Low-Dimensional Quantum Physics and Department of Physics, Tsinghua University, Beijing 100084, P.R. China;

[5] Frontier Science Center for Quantum Information, Beijing 100084, P. R. China;

[6] Collaborative Innovation Center of Quantum Matter, Beijing 100084, P.R. China.

*#These authors contributed equally to this work.*

*To whom correspondence should be addressed.*

*Emails:* H.L. liuhy@baqis.ac.cn and Q.X. qihua_xiong@tsinghua.edu.cn

# Abstract

Chirality, a pervasive phenomenon in nature, is widely studied across diverse fields including the origins of life, chemical catalysis, drug discovery, and physical optoelectronics. The investigations of natural chiral materials have been constrained by their intrinsically weak chiral effects. Recently, significant progress has been made in the fabrication and assembly of low-dimensional micro and nanoscale chiral materials and their architectures, leading to the discovery of novel optoelectronic phenomena such as circularly polarized light emission, spin and charge flip, advocating great potential for applications in quantum information, quantum computing, and biosensing. Despite these advancements, the fundamental mechanisms underlying the generation, propagation, and amplification of chirality in low-dimensional chiral materials and architectures remain largely unexplored. To tackle these challenges, we focus on employing ultrafast spectroscopy to investigate the dynamics of chirality evolution, with the aim of attaining a more profound understanding of the microscopic mechanisms governing chirality generation and amplification. This review thus provides a comprehensive overview of the chiral micro-/nano-materials, including two-dimensional transition metal dichalcogenides (TMDs), chiral halide perovskites, and chiral metasurfaces, with a particular emphasis on the physical mechanism. This review further explores the advancements made by ultrafast chiral spectroscopy research, thereby paving the way for innovative devices in chiral photonics and optoelectronics.

# 1. Introduction

The term "chirality" refers to an object that cannot be superimposed onto its mirror image through any combination of rotations and translations[1]. Since the discovery of optical rotation in chiral materials in the early 19[th] century[2], chiral spectroscopy methods based on circularly polarized light have become indispensable experimental techniques in chirality research[3]. In general, natural chiral materials exhibit relatively weak intrinsic chirality. Recent advancements in synthesis and assembly methods have facilitated the construction of low-dimensional nano-chiral materials and their hierarchical architectures, thereby significantly enhancing chiral optical activity and expanding the realm of research on chiral materials[4]. For instance, two-dimensional transition metal dichalcogenides (TMDs)[5-9] and their heterostructures[10-12] possess unique electronic band structures that exhibit chirality-dependent spin polarization coupling with electronic valleys, rendering them ideal candidates for valleytronics devices. Similarly, two-dimensional halide perovskites[13-15], possessing strong spin-dependent physical performance, particularly when incorporating certain chiral organic ligands, can impart pronounced chiroptical responses. Moreover, metallic[16] or dielectric metamaterials[17, 18] and metasurfaces[19, 20], composed of micro/nanostructured arrays, provide novel possibilities for the generation, manipulation, and enhancement of chiral optical effects. These engineered structures can break the symmetry and enable the design of chiral devices capable of selective interaction with circularly polarized light[21, 22]. The advancements in chiral materials not only enhance our comprehension of fundamental physics but also pave the way for technological innovations that have the potential to revolutionize various industries[23, 24].

The emergence of ultrashort pulsed lasers[25-27], distinguished by their monochromaticity, high brightness, and ultrashort pulse durations, has revolutionized the field of chiral spectroscopy. These lasers enable the advancement of spectroscopic methods with high temporal and spatial resolution, which are crucial for studying chiral materials, unveiling novel physical phenomena, and elucidating microscopic physical mechanisms. Circularly polarized light possesses inherent angular momentum, enabling the exchange of momentum with the spins of electrons, excitons, and other quasiparticles in chiral materials. This interaction enables the detection of the dynamical evolution of chiral optical activity, including processes such as electron and spin transfer, phonon scattering, and electron-phonon coupling. By gaining a profound comprehension of the micro mechanisms of chiral materials, researchers can devise novel strategies to achieve enhanced optical activity. The development of time-resolved methods employing intense laser has led to a remarkable enhancement in spin polarization, chirality, and nonlinear optical responses, which are previously unattainable through steady-state techniques. Clarifying the ultrafast dynamics governing the generation, transmission, and amplification of chiral mechanisms, as well as achieving precise multi-field control of chiral states, represent cutting-edge research frontiers that hold significant promise for design of chiral materials and their potential applications.

This review focuses on the ultrafast chirality-dependent dynamics of low-dimensional micro- and nano- chiral materials *via* helicity- and time-resolved spectroscopy techniques. We begin with an overview in the introduction, followed by a summary of the experimental methods and principles of pump-probe technology employing circularly polarized beams. Our focus is primarily directed towards elucidating the dynamics of chirality generation and transfer in TMD materials and 2D perovskites, as well as the ultrafast chiral response of artificial metasurfaces. Subsequently, we explore recent representative studies, including spin relaxation dynamics, spin-selective exciton and spin polarization dynamics. Lastly, we summarize further advancements in time-resolved optical spectroscopy, including research directions and potential applications.

# 2. Fundamentals and Experimental methods

## 2.1 Chirality and application

The phenomenon of chirality is pervasive in nature[1], characterized by objects that cannot be superimposed onto mirror images through rotations, translations, or inversions. The manifestation of their symmetry only occurs through reflection in a mirror, a property known as mirror reflection symmetry. This property can be likened to the distinctiveness between left and right hands (Fig. 1e), thus denoted by the term "chirality". As an inherent characteristic, it can be observed across a wide range of scales, encompassing the spin of electrons (Fig. 1a), circularly polarized light (Fig. 1b) and even the spiral structure of galaxies (Fig. 1f). Chirality plays a crucial role in biological systems[28]; numerous biological macromolecules, including DNA (Fig. 1d), the "codebook" of biodiversity on earth, and other nucleic acids, exhibit chirality[29]. The spatial configuration can significantly influence the properties and interactions of molecules, particularly in biological contexts where chiral molecules often exhibit different biological activities or reactivities. In the field of chemistry[30], chirality refers to molecules that possess identical constituents but exhibit distinct spatial arrangements, resulting in non-superimposable mirror images with different properties. The investigation of chiral materials holds immense significance across various scientific fields, with wide-range applications including material and optical research[31, 32].

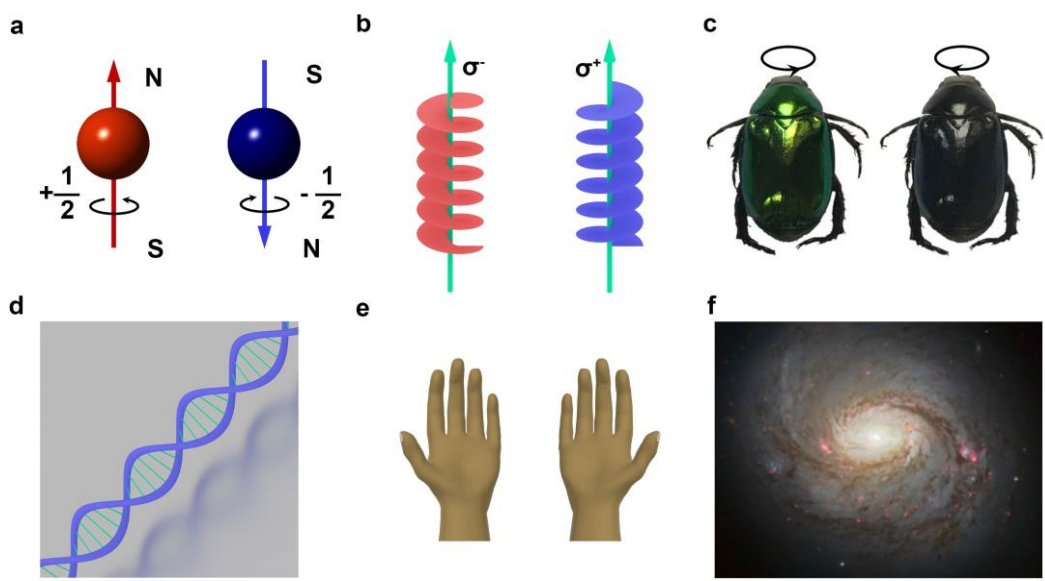

**Fig. 1** Chirality in nature. (a) Electrons with the opposite spins. (b) Left and right circular polarized light. (c) Photographing a scarab beetle (a type of beetle) using a left and right circular polarized light, respectively. Reproduced with permission from Ref. [33]. Copyright 2019, German Chemical Society. (d) The double helix structure of DNA. (e) Left and right hand. (f) A galaxy with a spiral structure. Reproduced with permission from NASA. Copyright 2013, NASA image galaxies.

Circularly polarized light (CPL), which enables control over both photon energy and angular momentum, is regarded as an exceptionally promising source for chirality studies via light-matter interactions. CPL carries angular momentum determined by its helicity, with $\sigma- = -\hbar$ and $\sigma+ = +\hbar$ in the propagation direction, representing left (LCP) and right circular polarization (RCP), respectively. The phenomenon of chirality transfer between light and matter enables detection, excitation or manipulation of chiral materials[34-38]. To evaluate the chirality of an object, one typically examines its differential response to electromagnetic fields with differing handedness, denoted as circular dichroism (CD) generally in the visible/near-infrared range. CD spectroscopy is an accessible analytical tool for investigating the inherent optical properties of chiral materials, arising from the disparity in extinction coefficients between LCP and RCP light. When LCP and RCP light propagate through a chiral medium, the intensity of each transmitted component exhibits disparity due to the differential absorption characteristics of LCP and RCP light. The resulting difference is traditionally quantified as the ellipticity, denoted by the ratio of the difference in intensities of LCP and RCP light to their sum, and expressed by $CD = (I_L - I_R)/(I_L + I_R)$. Therefore, the CD amplitude and spectral weight are directly related to the chirality of materials. However, the practical applications of natural chiral materials have historically been limited by their typically low CD signals, which pose challenges in terms of detection and utilization. Multiple techniques are realized by implementing not only CPL absorption but also CPL emission and scattering. These include helicity-resolved photoluminescence (PL) for the detection of PL polarization from excited states, vibrational circular dichroism (VCD) or Raman optical activity (ROA) spectroscopy for the analysis of circularly polarized intraband transitions, terahertz circular dichroism (TCD) spectroscopy to resolve chiral phonon modes in the long wavelength range, *etc*. Relevant in-depth information about these methods can be found in other excellent review articles[35, 39].

The emergence of low-dimensional materials and structures has presented opportunities for investigating chiral materials, leading to the exploration of numerous novel optical phenomena, such as circularly polarized light emission and spin-charge transfer. These phenomena hold considerable potential for applications in asymmetric catalysis[40, 41], chiral recognition[42, 43], chiral optical sensing[44, 45], disease treatment[46], chiral polarized light source[47] and chiral biomimetics[33, 35], as shown in Fig. 2. Despite these limitations, significant efforts have been made to enhance the CD signals of chiral materials, with strategies including supramolecular assembly[24] and surface-enhanced CD[20], thereby opening up new possibilities for their applications. Chiral photonic crystals and plasmonic chiral nanostructures, for instance, confine light at the nanoscale, significantly amplifying chiral effects. These structures possess the ability to selectively interact with circularly polarized light, thereby enabling applications in areas such as enantioselective sensing, optical switching, and the manipulation of light-matter interactions in chiral media. Understanding and harnessing chirality is essential in various scientific disciplines encompassing chemistry, biology, optics and materials science[28].

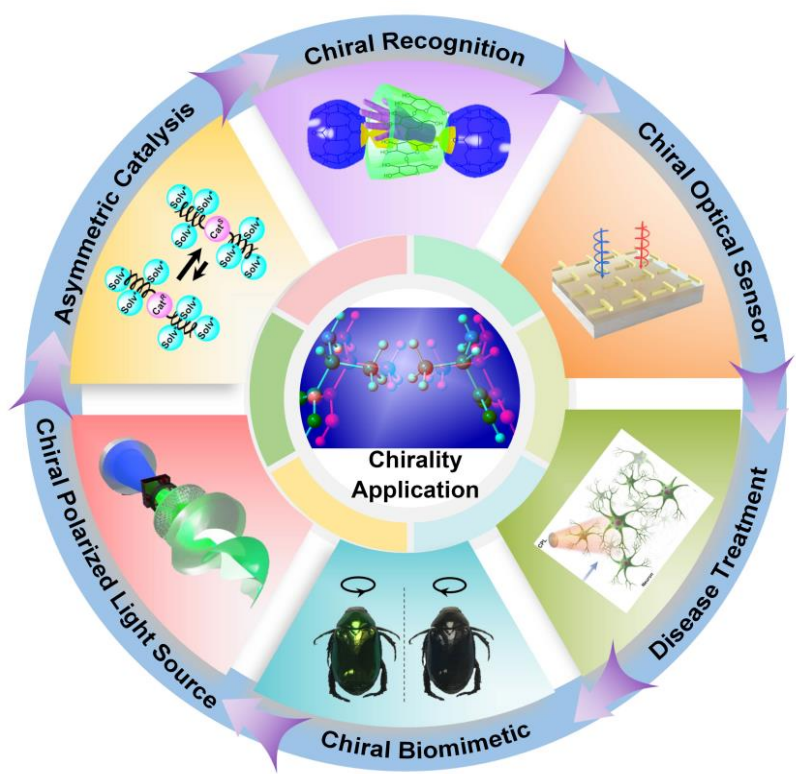

**Fig. 2** The application of chirality in numerous domains: asymmetric catalysis, Reproduced with permission from Ref. [41]. Copyright 2019, American Chemical Society; Chiral recognition, Reproduced with permission from Ref. [43]. Copyright 2018, Royal Society of Chemistry; chiral optical sensing, Reproduced with permission from Ref. [44]. Copyright 2023, American Chemical Society; disease treatment, Reproduced with permission from Ref. [46]. Copyright 2022, The authors; chiral polarized light source, Reproduced with permission from Ref. [47]. Copyright 2020, Springer Nature Limited; chiral biomimetic, Reproduced with permission from Ref. [33]. Copyright 2019, German Chemical Society.

## 2.2 Helicity-resolved transient absorption spectroscopy

Transient absorption spectroscopy (TAS), a robust method based on ultrafast light beams as pump and probe, is a powerful approach for investigating carrier dynamics[48-52] and many-body interactions[53, 54]. With the aid of this experimental method, it becomes feasible to investigate the temporal evolution of single-particle and collective excitations spanning from femtosecond to nanosecond timescale ($10^{-15}$ to $10^{-9}$ s). For instance, one can explore ultrafast processes such as carrier-carrier scattering, carrier-phonon interaction, as well as couplings among different degrees of freedom and order parameters in distinct time domains. The ultrafast dynamics in chiral materials encompass a wide range of chirality-related phenomena that can be reflected in the helicity of pump-probe beams, from the initial excitation to the subsequent exchange/relaxation processes occurring on ultrafast timescales. Therefore, helicity-resolved TAS, employing circularly polarized beams in TAS, is essential for the investigation of ultrafast chirality-dependent dynamics. In this section, we will provide a concise overview of the experimental setup for helicity-resolved TAS, elucidate its principle and outline the methodology for data analysis. For those interested in the origins and evolution of this technique, we recommend readers refer to these excellent reviews[51, 53, 55, 56], which offer a comprehensive guide to this field, ranging from its pioneering experiments to state-of-the-art applications in material science, chemistry, and biology.

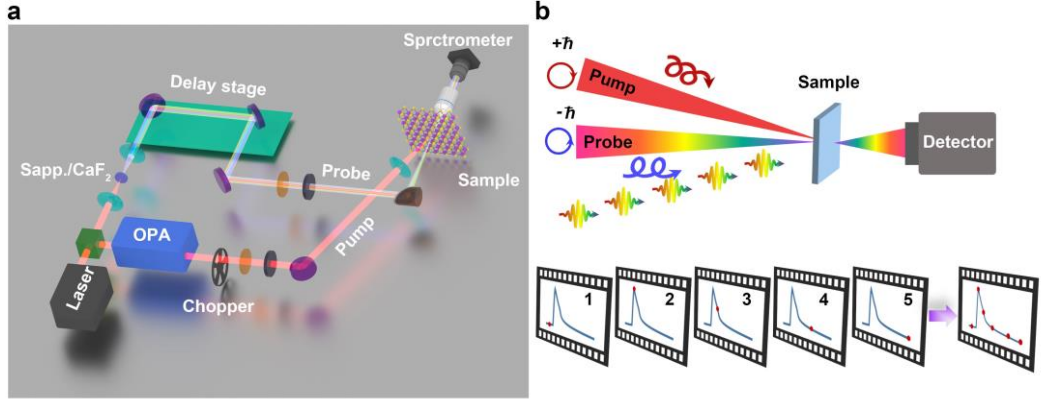

**Fig. 3** Helicity-resolved TAS. (a) The optical setup of helicity-resolved TAS. 800 nm laser is generated from the laser amplifier system, and splitted into two beams as pump and probe. The pump pulse is more intense to excite the sample, and probe pulse used here is to monitor the evolution of dynamics at the different delay time. (b) The schematic of helicity-resolved TAS with circularly polarized pump-probe.

Fig. 3a schematically illustrates the typical optical setup of helicity-resolved TAS. The pump-probe methodology involves two laser beams derived from the same laser source, with a center wavelength of 800 nm and pulse duration at few tens of fs. A circularly polarized pump with high fluence excites the sample from its equilibrium to a non-equilibrium state. Subsequently, a weaker probe beam with the same/opposite polarization captures snapshots of the temporal evolution of specific order parameters within the system as a function of time delay, as shown Fig. 3b. An optical parameter amplifier (OPA) is used to generate tunable pump from visible to near-infrared. The broadband probe involves the utilization of white-light generation (WLG) through nonlinear optics by focusing the 800 nm laser into sapphire or CaF$_2$ crystals. The pump-probe delay time is achieved through changing the optical path difference between two beams by moving a motorized delay stage. The polarizations of the pump and probe are controlled through a combination of a polarizer and wave plates ($\lambda$/2 and $\lambda$/4).

The probe beam, subsequent to its interaction with the sample, is collected by an optical spectrometer, and the observed change exhibits a close correlation with the disparity between equilibrium and non-equilibrium states. Taking a two-level system as an example, when the energy of the pump pulse exceeds the bandgap, electrons from the valence band (VB) undergo excitation and transiently accumulate at the bottom of the conduction band (CB). The probe pulse arriving at the sample at this moment is expected to exhibit a higher degree of transmission ($\Delta T/T > 0$) and lower absorption ($\Delta A < 0$), because the excited electrons hinder further excitation from VB to CB due to Pauli blocking[57]. Therefore, the change in $\Delta T/T$ or $\Delta A$ can be simply treated to be proportional to the population of quasiparticles in nonequilibrium $n_{pe}$, as expressed in equation 2-1. Generally, TAS can be performed by measuring either reflection or transmission, depending on the optical properties of the sample or substrate. Here we focus on the transmission mode as an example, disregarding reflection which is usually at least an order of magnitude lower than that of transmission for micro-/nano-materials on transparent substrates. The relation between absorption and transmission can be written as

$$\Delta T/T = 10^{-\Delta A} - 1 = \frac{I(\lambda)_{pump\_on} - I(\lambda)_{pump\_off}}{I(\lambda)_{pump\_off}} \propto n_{pe} \qquad (2\text{-}1)$$

where $I(\lambda)_{pump\_on}$ and $I(\lambda)_{pump\_off}$ represent transmitted probe spectra with pump pulse on and off, respectively. It can be simply written as $\Delta T/T \approx -2.3\Delta A$ by using Taylor series expansion and ignoring high orders due to the absorption change $\Delta A$ being on the order of $10^{-2}$ to $10^{-4}$. Analogously, in the case of opaque bulk materials or thin films on non-transparent substrates, the transient reflectivity change $\Delta R/R$ can be employed to investigate ultrafast dynamics[58, 59].

Now we turn to infer the ultrafast dynamics reflected from the helicity-resolved TAS data, by combining Fig. 3 and 4. Excited carriers undergo a series of relaxation processes before ultimately returning to the equilibrium state. From the perspective of helicity-resolved TAS, multiple analysis approaches can be employed:

i. Ground state bleaching (GSB): for over-gap excitation when the pump pulse energizes electron transition from VB to CB, leaving behind hole formation in VB. This induces a decrease in absorption due to the Pauli blocking effect which prevents further excitation of electrons from VB to CB. As a result, $\Delta T/T$ exhibits a positive peak while $\Delta A$ presents a negative dip signal.

ii. Photon induced absorption (PIA): following the pump excitation, there is a possibility for electrons in CB to undergo further excitation to much higher levels by absorbing additional photons from the probe. Consequently, this leads to an additional $\Delta A$ peak, corresponding to a negative dip in $\Delta T/T$.

iii. Blue/red shift: band gap renormalization, arising from the coherent interaction between light and matter, carrier-carrier interaction, *etc.*, can eventually fine-tune the absorption peak to either a blue or red shift, resulting in symmetric peak/dip features in $\Delta A$ and $\Delta T/T$.

iv. Helicity-resolved TAS: following the excitation of circularly polarized pump pulse, the probe pulse with the same circular polarization (SCP) detects quasiparticle dynamics possessing the same angular momentum as the pump. Conversely, an opposite circularly polarized (OCP) probe tracks population and relaxation with opposite angular momentum. This method enables the excitation and monitoring of ultrafast chiral dynamics, which are further elaborated in detail in the main text.

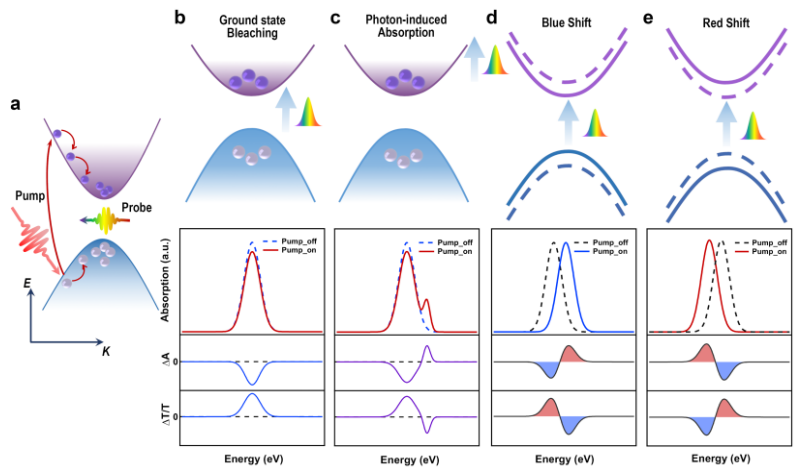

**Fig. 4** Photo-induced ultrafast dynamics. (a) The schematic of photo-excitation and detection. Electrons are excited from VB to CB, leaving the holes in VB. (b-e) The simulated transient signatures in material, including (b) Ground state bleaching (GSB), (c) Photon induced absorption (PIA), and (d, e) Band gap renormalization with blue/red shift.

# 3. Transition metal dichalcogenides

## 3.1 Crystal and electronic structure

Atomically thin TMDs, a prominent member of two-dimensional (2D) materials family, have garnered significant attention owing to their exceptional physical properties and potential applications in optoelectronics[60, 61] and valleytronics[5, 9, 62-64]. These materials exhibit a general form $MX_2$[65], where M is a transition metal atom (Mo, W, Ta, *etc*.) and X is a chalcogen element (S, Se, Te, *etc*.). The layers consist of an X-M-X sandwich configuration (see Fig. 5a upper panel), in which two chalcogen atoms X encapsulate tightly bonded transition metal atoms M. Although the covalent bonding within each layer is strong, the interlayer interaction is mainly governed by Van der Waals forces. Therefore, it can be readily reduced to few-layer even monolayer (ML) through the mechanical exfoliation technique, commonly used for graphene[66, 67]. The most common structures of TMD are trigonal prismatic (2*H*) and octahedral (1*T*) phases, which exhibit distinct stacking orders between transition metal and chalcogen atoms. In natural state, the 2*H* phase being more thermally stable than the 1*T* phase, is widely considered as the primary choice in many studies. Indeed, investigations of their bulk materials have been conducted since the early 1960s, with notable contributions by Wilson and Yoffe[68]. However, it is only in recent years that those propensities have gained widespread recognition and becomes a subject of extensive study due to the emergence of fabrication techniques mentioned above. For a comprehensive understanding of the electronic structure and research advancements in 2D TMD materials, interested readers can refer to the early review[60, 69].

In contrast to graphene with a Dirac-cone electronic structure, TMD ML materials are semiconductors with direct visible/near-infrared bandgap[62], characterize by strong photoluminescence spectroscopy (PL)[70], and high quantum yield[71], nearly four orders of magnitude higher than their bulk counterparts. As the material dimensions are reduced, the dielectric screening weakens from bulk to monolayer[72, 73], leading to the formation of excitons with large oscillator strength[74-76] and binding energy up to hundreds of meV[72, 77-81]. Fig. 5b shows the reflectance contrast derivative of $WS_2$ ML at 5 K, where clear Rydberg states $n = 1s, 2s, 2p...$ are observed. Notably, the intensity of the 1s states surpasses that of other states, thereby dominating optical transition and transport properties[82]. Experiments manifest the large exciton effects in strong and sharp resonance in PL[71], absorption/reflection[72] and photoconductivity[83] spectroscopy. Particularly, in intrinsic/artificial doped materials, the robust neutral exciton can attract extra electron or hole or even exciton to form a many-body complexes, known as trions (charged complexes)[84-86] or bi-excitons (neutral complexes)[87]. The binding energy of trions is roughly one-tenth that of neutral excitons, approximately 20 to 40 meV for different materials[84, 88, 89]. This difference arises from the additional charge carriers, which leads to extra Coulomb interactions in the many-body complexes. The unique properties of TMD MLs make it an ideal platform for studying exciton physics[74], optoelectronics[62, 90, 91] and light-mater strong coupling[92].

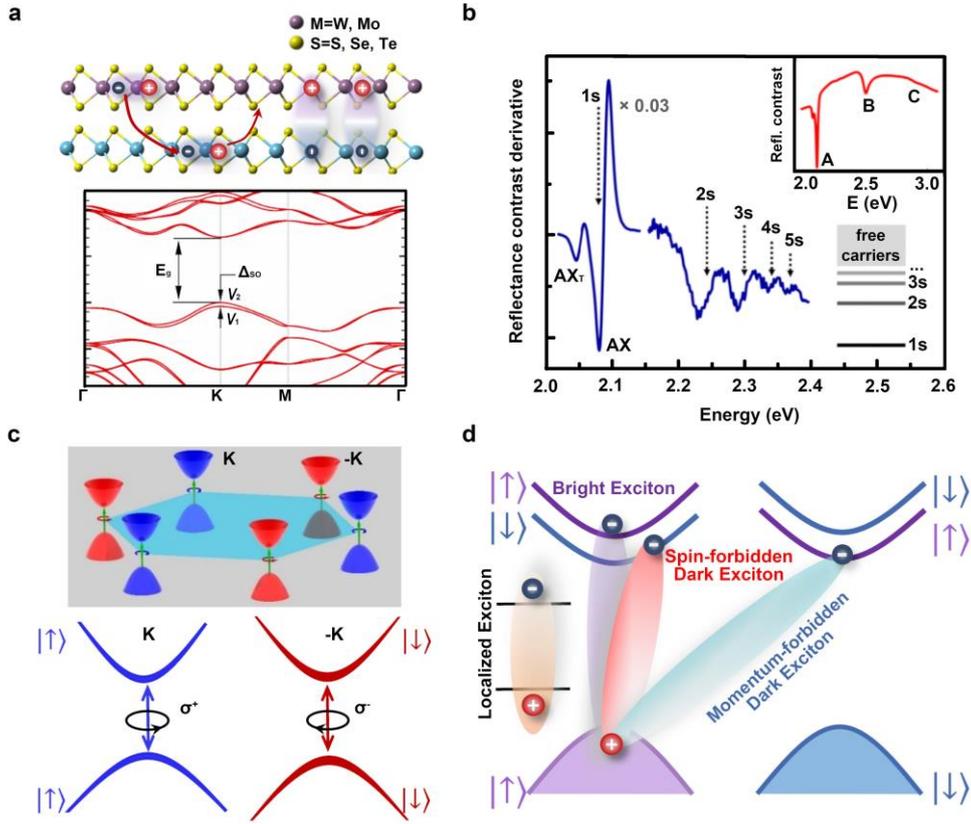

**Fig. 5** Lattice and electronic structures of TMD materials. (a) Upper panel: TMD crystal structure with a general form $MX_2$, where M is a transition metal atom (Mo, W, Ta, *etc*.) and X is a chalcogen element (S, Se, Te, *etc*.). The layers consist of an X-M-X sandwich configuration, in which two X atoms encapsulate a tightly bonded M. It can form intralayer and interlayer with the electron and hole reside in different layer. Lower panel: band structure of TMD ML, exhibiting larger spin splitting in VB at K valley. Reproduced with permission from Ref. [74]. Copyright 2012, American Physical Society. (b) Exciton Rydberg states in the derivative of the reflectance contrast spectrum of the $WS_2$ monolayer at 5 K, the Rydberg states $n = 1s$, $2s$, $2p...$ are clearly observed. The inset shows the energy positions of A, B, and C excitons, respectively. Reproduced with permission from Ref. [72]. Copyright 2014, American Physical Society. (c) K and -K (K') valleys in the corner of Brillouin zone marked by blue and red color respectively. The lower part illustrates the optical selection rule, in which can be selectively excited by left- and right circular polarized light. Reproduced with permission from Ref. [5]. Copyright 2014, Springer Nature Limited. (d) Bright and dark excitons in TMD materials.

The electronic band structures of TMD materials exhibit a hexagonal pattern[93], wherein VB and CB possess extrema located at the corners of the Brillouin zone, known as K valleys analogous to those of graphene[94]. It is worth mentioning that valley properties strongly influence optical absorption, carrier mobility and effective mass in TMDs. For TMD MLs, the broken symmetry of spatial inversion leads to an intriguing scenario where the energetically degenerate K and -K (K') valleys[95] are inequivalent in momentum space (Fig. 5c). A large spin splitting of few hundreds of meV emerges in the VB due to strong spin-orbit coupling[96, 97] originating from the partially filled *d*-orbitals of the heavy metal atoms[98], as shown in Fig. 5a lower panel. In the meanwhile, there is a relatively smaller CB spin splitting due to partial compensation between *p*- and *d*-state contributions[96-98]. In TMD MLs, the VB spin splitting at the K valley is approximately 0.2 eV for molybdenum (Mo)-based TMDs[74, 99-101] and 0.4 eV for tungsten (W)-based TMDs[102, 103], giving rise to two sub-bands that correspond to two types of excitons with different energy levels, designated A and B. As a consequence of spin-orbit coupling and the breaking of inversion symmetry, electrons in adjacent valleys hold opposite angular momentum, known as spin-valley locking. As such, valley-dependent optical selection rules govern physical processes at a fundamental level, as illustrated in Fig. 5c lower panel. Right circularly polarized light (σ+) can excite electrons in the K valley, resulting exciton PL that conserves the same helicity with incident light. Conversely, left circularly polarized light (σ-) allows excitation of electrons in the -K (K') valley[5, 6, 8, 9, 104, 105]. Besides these bright states that satisfy the optical selection rules, there may exist spin- and momentum-forbidden dark exciton states with different energies, depending on the exchange and spin–orbit interactions[106], thereby producing rich exciton species, as seen in Fig. 5d. The spin-valley coupling offers the possibility of manipulating the valley degree of freedom, thereby enabling applications of quantum and classical processing devices and playing a crucial role in the field of valleytronics[64, 107, 108].

## 3.2 Valley depolarization dynamics in TMD MLs

The manipulation of valley degree of freedom constitutes a pivotal focus in valleytronics applications, aiming to exploit the valley polarization for quantum computing and information processing[9, 108, 109]. Valley depolarization refers to the gradual decay of quantum coherence between valley quantum states over time, which can be caused by various mechanisms. The most common scenario is chirality-dependence depolarization due to long- and short-range interactions including intravalley scattering[110-112], electron-electron interaction[113, 114], and Coulomb-induced[115-117] or photon-mediated intervalley scattering[110]. Fig. 6a shows the schematic of helicity-resolved TAS that allows to track ultrafast valley-depolarization dynamics. In the case of resonant or near-resonant conditions in TMD MLs, circularly polarized light selectively pumps excitons within the K or -K (K') valley, following spin-valley locking and optical selection rules. Then the valley depolarization evolution is monitored by measuring the sample transmission/reflection using same/opposite circularly polarized probe pulses (SCP/OCP). The difference in TAS signals between the SCP and OCP configurations is closely related to the valley depolarization dynamics. Fig. 6b shows the valley relaxation dynamics in 1L-WS$_2$ by Zilong Wang[110] *et al.* They found that upon A excitons injection in the K valley by $\sigma+$ resonant pump, an almost instantaneous buildup of B exciton in K valley occurs with a short time constant of ~200 fs at 77 K. This process is highly temperature-dependent, suggesting a phonon-assisted relaxation depending on the value of the CB spin splitting. Theoretical calculations demonstrated that intravalley spin−flip scattering occurs only at the K point on longer time scales, while the occupation of states located away from the CB minimum significantly reduces the scattering time. Indeed, the contribution of intravalley scattering event to the depletion of PIA process was found to be minimal, thus the main decay channel can be reasonably attributed to intervalley interaction[112]. As shown in Fig. 6d, the data illustrates the differential transmission signal $\Delta T/T$ of MoS$_2$ under excitation of SCP and OCP at 74 K. Notably, the difference between these two cases gradually diminishes, as the time delay varies from 0 to 15 ps, suggesting a significant influence of strong Coulomb interactions on valley population dynamics[116]. These bright excitons coherently couple to dark intervalley excitonic states, which facilitate fast electron valley depolarization (Fig. 6c). Upon $\sigma+$ pump excitation, bright excitons are initially formed at K valley, followed by rapid electron scattering to the -K driven by Coulomb interactions. Next, the holes located in K valley coherently couple to the -K valley through spin-flip process, resulting in a delayed relaxation time of about 10 ps due to nondegeneracy of the VB spin states[115]. This also explains the lower valley polarization in the steady-state PL spectrum. R Schmidt *et al.* demonstrated the Coulomb-induced intervalley depolarization in atomically thin WS$_2$ through a comprehensive combination experimental and theoretical investigations[117]. They also found a band gap renormalization of the pumped K valley and a red-shift of excitonic resonance of the unpumped -K valley. Additionally, zone-edge phonon-assistant carrier scattering event also plays an important role in the process of valley-depolarization. Soungmin[118] *et al.* reported a definitive determination of the longitudinal acoustic (LA) phonons at the K point, which are responsible for the ultrafast valley depolarization in monolayer MoSe$_2$, as shown in Fig. 6e. They used sub-10-fs resolution pump-probe spectroscopy to measure the differential transmission of sample, revealing coherent phonons signals with peaks at 4.65 and 7.37 THz, corresponding to the first-order LA mode and optical $A_1^{'}(\Gamma)$ mode, respectively. Fig. 6f shows both simulated and experimental $\Delta T(t)/T_0$ of $A_1^{'}(\Gamma)$, LA(K) mode and combination of $A_1^{'}(\Gamma)$ and LA(K) mode. In conjunction with phonon-symmetry analysis and first-principles calculations, they proposed that the dominant mechanism for acoustic LA(K) phonon generation is intervalley scattering, rather than the impulsive stimulated Raman scattering, which primarily contributes to the generation of the A1′(Γ) mode. Recently, X. Bai *et al.* observed chirality-dependent trion dynamic closely depending on temperature in monolayer WS$_2$ by helicity-resolved TAS with pump photon tuned at the trion resonance[119]. This unveils phonon-assisted up-conversion process and depolarization process. Other ultrafast technology techniques, such as time-resolved Kerr rotation and transient Faraday rotation, serve as alternative and highly sensitive tools to studying spin dynamics, with more detailed information refer to paper[120]. For instance, two distinct time scales ($\tau_1 = 200$ fs and $\tau_2 = 5$ ps) for valley degenerate relaxation, have been found by transient Faraday rotation in MoS$_2$[121]. The fast decay $\tau_1$ is attributed to the intervalley electron-hole Coulomb exchange (similar results have also been reported in WSe$_2$[122]), and the slow one $\tau_2$ is primarily due to long-range interactions[112, 123].

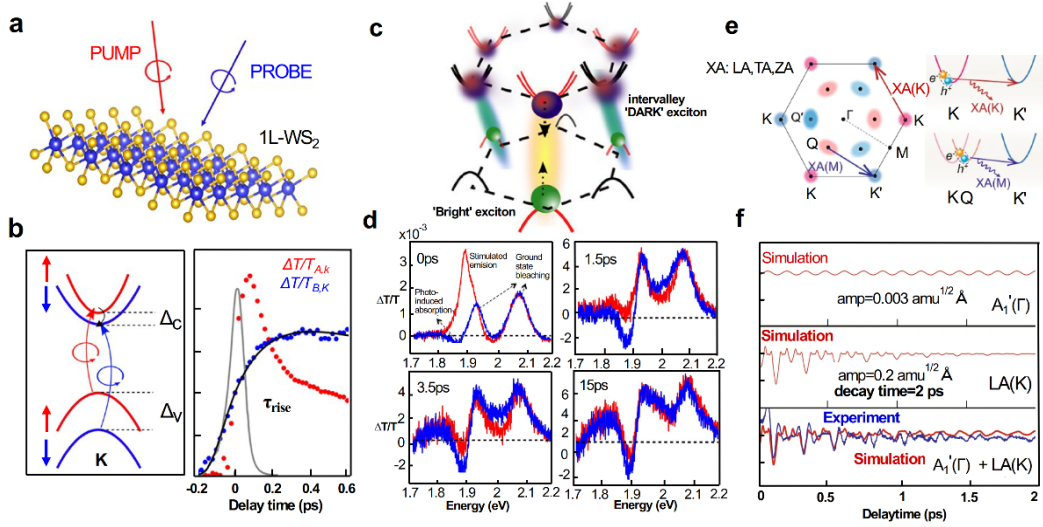

**Fig. 6** (a-b) The schematic of helicity-resolved transient absorption measurement, A excitons are injected in the K valley by a circularly polarized pump (red arrow). The probe pulse has the same helicity, and it is resonant with the B exciton (blue arrow), red and blue curves are $\Delta T/T$ at 77 K around the A and B transitions. Reproduced with permission from Ref. [110]. Copyright 2018, American Chemical Society. (c) The intervalley bright exciton transition and bright-to-dark exciton transition processes. (d) Transient absorption spectra of monolayer $MoS_2$ at 74 K with various delays between pump and probe beams. SCP (red line) and OCP (blue line). (c) and (d) Reproduced with permission from Ref. [116]. Copyright 2014, American Chemical Society. (e) Valley scattering processes mediated by XA(K) and XA(M) phonons (XA: LA, TA, and ZA). LA：Longitudinal acoustic phonons TA：Transverse acoustic phonons and ZA: out-of-plane acoustic phonon mode, which is sometimes called the "zigzag" mode. (f) Simulated differential transmittance $\Delta T(t)/T_0$ of the $A_1'(\Gamma)$ mode (upper panel), LA(K) mode (middle panel) and combination of the LA(K) and $A_1'(\Gamma)$ modes (bottom panel). (e-f) Reproduced with permission from Ref. [118]. Copyright 2022, Springer Nature Limited.

## 3.3 Valley-dependent optical Stark effect in TMD MLs

The optical Stark effect (OSE) is a phenomenon of band shift/splitting that arises from the coherent interaction between electric field of light and materials, providing a means to modify and manipulate electronic structure and spin/pseudospin in quantum systems, such as atomic quantum gases[124], quantum dots[125], and III–V quantum wells[126]. The OSE can be understood from Floquet theory[127], which describes the quasi-static eigenstates driven by a periodic electromagnetic field within the system. A simple example is given in Fig. 7a upper panel, illustrating a two-level system with the ground state and the exciton state under the illumination of monochromatic light. This can be described using a semi-classical Hamiltonian[128], written as

$$\hat{H}(t) = \hat{H}(0) + \hat{V}(t) \tag{3-1}$$

where $\hat{H}(0)$ represents the equilibrium Hamiltonian, which describes the two-level eigenstates $|a\rangle$ and $|b\rangle$. $\hat{V}(t)$ is the perturbation term in system, which is proportional to the electric field of the light $\varepsilon(t)$. In the dipole approximation, the perturbation term can be expressed as

$$\hat{V}(t) = \hat{p}\varepsilon(t)$$
$$\varepsilon(t) = \varepsilon_0 \cos(2\pi\nu t) \tag{3-2}$$

$\hat{p}$ is the electric dipole moment operator of atom, $\varepsilon_0$ and $\nu$ are amplitude and frequency of the oscillating electric field associated with light. The perturbation term interacts with equilibrium states $|a\rangle$ and $|b\rangle$ in the form of light absorption/emission, creating transient Floquet (photon-dressed) states $|a+h\nu\rangle$ and $|b-h\nu\rangle$ (Fig. 7a upper panel, higher order terms are not presented here). The Floquet bands can hybridize with equilibrium states, resulting in energy repulsion between the Floquet and equilibrium states. This picture is analogous to the Coulomb interaction-induced hybridization between two atomic orbitals, resulting in the formation of bonding and anti-bonding orbitals[129]. In this case, the energy repulsion induces an energy shift of the optical transition between $|a\rangle$ and $|b\rangle$, which can be described by the equation[130]

$$\Delta E = \frac{M_{ab}^2 \langle \varepsilon^2 \rangle}{\Delta} \tag{3-3}$$

where $M_{ab}$ is the polarization matrix element between the two equilibrium states $|a\rangle$ and $|b\rangle$, $\langle \varepsilon^2 \rangle$ is the time average of the electric field square (equals to $\varepsilon_0^2 / 2$) that is proportional to the light intensity, and $\Delta$ is the laser detuning energy defined as $\Delta = E_b - E_a - E_{pump}$.

Elbert J. Sie $et\ al.$ employed helicity-resolved TAS with pump energy below exciton resonance to study the valley-dependent OSE dynamics in monolayer WS$_2$[128]. As shown in the lower panel of Fig. 7a, a sharp peak of $\Delta\alpha$ (absorption change) is observed during the pulse duration ($\Delta t = 0$ ps) for the SCP ($\sigma$- pump and $\sigma$- probe) configuration, while no discernible signal is detected in the OCP ($\sigma$- pump and $\sigma$+ probe) case. The authors estimated a selective tuning of the exciton level tuned by up to 18 meV, providing a clear demonstration of broken valley degeneracy. Kim $et\ al.$ reported the valley-selective OSE in WSe$_2$ ML using a non-resonant pump, demonstrating a chirality-dependent energy shift of more than 10 meV. This is equivalent to a valley pseudo-magnetic field as large as $\sim$60 T in equilibrium[131]. As depicted in Fig. 7b, the coupling between the photon-dressed ground state and A-exciton at K valley results in energy-level repulsion and a strong transient OSE signal at A-exciton resonance when probed by the $\sigma$+ light. On the contrary, A exciton at -K (K') valley remains unaffected by the $\sigma$+ pump due to the optical selection rules, resulting in no production of transient signals by the $\sigma$- probe. Yong $et\ al.$ focused on monolayer MoSe$_2$, and further demonstrated the coupling of the driving light field with multiple exciton states, when the inter-valley biexciton states play a dominant role in the OSE[132]. Specifically, the exciton-biexciton coupling breaks the valley selection rules based on non-interacting excitons as shown in Fig. 7c. The upper panel of Fig. 7c illustrates the optical transition and selection rules for exciton and many-body biexciton states. This picture determines the anomalous behavior with manifestation of the many-body OSE, where driving photon couples to different excitonic states. The lower panel of Fig. 7c displays the energy shift induced by OSE as a function of detuning energy $|\Omega_X|$ ($\Omega_X \equiv E_p - E_X$, $E_p$ is the pump photon energy and $E_X$ is the A exciton energy), and change from redshift at large $|\Omega_X|$ to blueshift at small $|\Omega_X|$. This OSE is described by $\delta E_K = \frac{V_b^2}{|\Omega_X|} - \frac{V_b^2}{|\Omega_X| - E_b}$, where $V_b^2$ is the exciton-photon coupling strength and $E_b$ is the biexciton binding energy. The first term arises from the energy shift of ground state, and the second term comes from the coupling between exciton and biexciton state[132]. They also determined the binding energy of the inter-valley biexciton at 21 meV, and the transition dipole moment of the exciton-biexciton transition at 9.3 Debye. This study reveals the key role of many-body effects in the coherent light-matter interaction in atomically thin two-dimensional materials. These results provide strong support for the utilization of helicity-resolved TAS as a powerful tool to investigate the valley-dependent OSE and realize Floquet engineering of excitons.

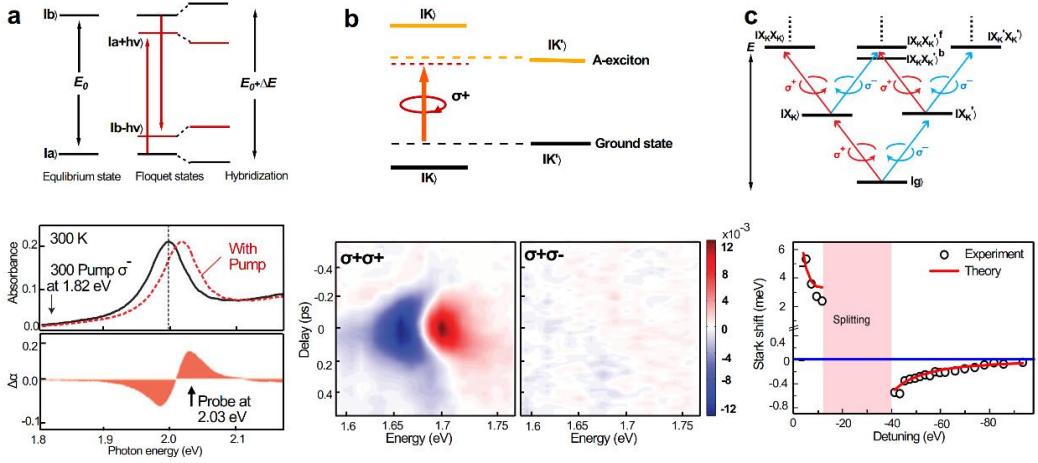

**Fig. 7** (a) Upper panel: Energy level diagram of two-level $|a\rangle$ and $|b\rangle$ atoms showing that the equilibrium and photo-dressed state $|a+h\nu\rangle$ and $|b-h\nu\rangle$, resulting in energy shifted after hybridization via optical Start effect in WS$_2$. Lower panel: Measured absorbance of monolayer WS$_2$ (black) and a hypothetical absorbance curve (dashed) that simulates the optical Stark effect. Reproduced with permission from Ref. [128]. Copyright 2014, Springer Nature Limited (b) Upper panel: Schematic of the optical Stark effect in WSe$_2$ for valley transitions with nonresonant, left circularly polarized σ+ pump. The dashed black and yellow lines denote the unperturbed ground and exciton states, respectively. Lower panel: Transient reflection spectra of A-exciton resonance at 77 K. with probe helicity same and opposite from pump, respectively. Reproduced with permission from Ref. [131]. Copyright 2014, Springer Nature Limited. (c) Upper panel: An illustration of optical transition and selection rules for excitons and many-body biexciton states. $|g\rangle$, $|X_k\rangle$ ($|X_{k'}\rangle$) and $|X_kX_k\rangle$ ($|X_kX_{k'}\rangle$) denote the ground state, K (K′) valley exciton state and intra-/inter- valley biexciton states. Lower panel: Optical Stark shift of the K′-valley exciton is well defined away from the exciton–biexciton resonance, and it changes from a redshift at large $|\Omega_X|$ to a blueshift at small $|\Omega_X|$. The experimental data (black circles) agree quantitatively with the prediction of the theoretical model (red lines). Reproduced with permission from Ref. [132]. Copyright 2018, Springer Nature Limited.

# 4. Halide Perovskites

Metal halide perovskites and perovskite-based devices have gained significant attention in recent years due to their exceptional optical and electronic properties. As a direct band-gap semiconductor, they exhibit great potential for applications of solar-cell materials[133-135], photodetectors[136], photonic lasers[137, 138] and light emitting devices[139, 140]. Helicity-resolved TAS is capable of revealing chiral dynamics in perovskites originating from spin-dependent quantum effects, such as spin-flip dynamics, optical Stark effect, and magneto-optical effect. These phenomena are attributed to giant spin-orbit coupling[141, 142], strong electron-phonon interaction[143, 144], and inherent inversion asymmetry[142]. A comprehensive understanding of spin-dependent physics in perovskites serves as a fundamental basis for material design and the advancement of spintronics applications.

## 4.1 Crystal and electronic structures of halide perovskites

Generally, perovskites refer to crystals with ABX$_3$ structure, where A is a monovalent cation (such as alkali cations or methylammonium MA$^+$ cations), B is a bivalent cation (such as Pb$^{2+}$ or Sn$^{2+}$), X is a monovalent anion (such as halide anion or mixed halide anions), as shown in Fig. 8a. Typical halide perovskite compounds include inorganic CsPbX$_3$[145] and organic-inorganic MAPbX$_3$[146]. Among them different A cations lead to structural variations (e.g., CsPbBr$_3$ is better stable in the orthorhombic phase at room temperature[147], while MAPbBr$_3$ adopts the cubic phase [148]) and changes in optical band gap (MAPbBr$_3$ has a lower band gap[149]). Through compositional tuning and structural engineering, the optical band gap in perovskites can be continuously controlled from the mid-infrared to the visible range, and the exciton binding energy can be increased from few to several hundreds of meV[150-152]. For instance, strong multi-color emission occurs at room temperatures in cesium lead halide perovskites (CsPbX$_3$, X = Cl, Br or I) by varying the halide composition[153, 154]. Similar to other 2D materials, the reduction of perovskite thickness down to monolayers significantly modifies the electronic and optical properties, including bandgap broadening[155], third-harmonic generation[156], enhanced light absorbance[157], etc. These grant the possibility to fabricate heterostructures and artificial superlattices[158] and hold immense promise for the future development of innovative devices[159, 160].

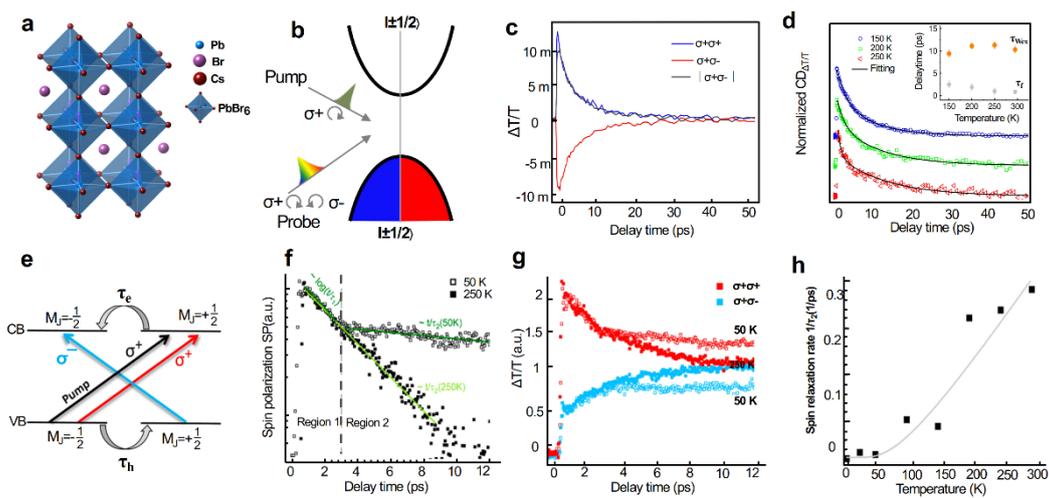

**Fig. 8** Crystal structure band structure and spin selection rules in perovskites. (a) Orthorhombic structure of CsPbBr$_3$. (b) Schematics of the electronic band structure of CsPbBr$_3$ and pump/probe exciton states with total angular momentum ±1 with circular polarization ($\sigma$+/$\sigma$-). The double degeneracy of spin states at the CBM and VBM have angular momentum ±1/2. (c) Ultrafast exciton dynamics of CsPbBr$_3$ with pump-probe of the same ($\sigma$+$\sigma$+) or opposite-polarization ($\sigma$+$\sigma$-). (d) Normalized signal of the circular dichroism with vertical offsets as a function of lattice temperature at a fixed pump fluence. (b-d) Reproduced with permission from Ref. [161]. Copyright 2020, Springer Nature Limited. (e) Optical transitions between VB and CB excited by circularly polarized light. (f) Spin polarization dynamics of CsPbI$_3$ nanocrystal at 50 and 250 K plotted on a logarithmic scale. (g) Transient dynamics of CsPbI$_3$ nanocrystal at 50 and 250 K with different pump-probe combinations. (h) Spin relaxation rate $1/\tau_2$ versus temperature. (e-h) Reproduced with permission from Ref. [162]. Copyright 2020, American Chemical Society.

## 4.2 Ultrafast spin-flip dynamics in perovskites

The investigation of spin dynamics, particularly determination of the spin relaxation or decoherence time in perovskites, plays a crucial role for the realization of spintronic devices. Taking CsPbBr$_3$ as an example, the conduction band minimum (CBM) is related to Pb ($6p$) orbitals, whereas the valence band maximum (VBM) comes from a hybridization of Pb ($6s$) and Br ($4p$) orbitals[142, 163]. Because of strong spin-orbit coupling, these bands are mixed with spin, giving rise to double degenerency with electrons and holes characterized by angular momentum $j_{e(h),z} = \pm 1/2$ [142, 161, 164], as shown in Fig. 8b. Consequently, spin-dependent selection rules, i.e., carriers/excitons with angular momentum $J_z = j_{e,z} + j_{h,z} = \pm 1$ exhibit spin-allowed or spin-forbidden upon interaction with circularly polarized $\sigma+/\sigma-$ light. Zhao *et al.* employed helicity-resolved TAS with both SCP and OCP ($\sigma+\sigma+$ and $\sigma+\sigma-$) pump-probe configurations to investigate the photoinduced circular dichroism and exciton spin dynamics in CsPbBr$_3$ films[161]. Following the injection +1 excitons by $\sigma+$ pump light, the decay/growth of +1/-1 excitons can be monitored by $\sigma+/\sigma-$ probe light. As shown in Fig. 8c, +1/-1 excitons exhibit symmetrical signal and merge at ~50 ps, with the exception of a minor fast decay of +1 exctions in the first few ps. Zhao *et al.* proposed an electron-hole exchange interaction that flips +1 excitons into -1 excitons, giving rise to transient circular dichroism in the kinetics. The transient circular dichroism in CsPbBr$_3$ (defined as $\Delta T/T_{\sigma+\sigma+} - \Delta T/T_{\sigma+\sigma-}$) does not show noticeable temperature dependence (Fig. 8d), in agreement with the Coulomb interaction that stays constant across varying lattice temperatures[165]. The electron-hole exchange interaction can be described by the Bir-Aronov-Pikus (BAP) model[166], for the scenario of electron spin relaxation exchange via scattering on holes, written as[167]

$$\frac{1}{\tau_s} \propto N_p \alpha_B^3 \frac{\Delta^2}{E_B} k \alpha_B$$

where $\tau_s$ is the spin relaxation, $N_p$ is hole concentration, $\alpha_B$ is exciton Bohr radius, $\Delta$ is the exchange coupling, and $E_B$ is the exciton binding energy. In BAP model, electron-hole exchange interaction flips exciton constituents simultaneously, resulting in an excellent balance of robust exciton resonances, optical nonlinearities and relatively long exciton spin lifetimes, which are consistent with experimental findings in Ref. [161]. In a study of 2D (C$_6$H$_5$C$_2$H$_4$NH$_3$)$_2$PbI$_4$ thin films, the exciton spin relaxation was also ascribed to the control of the BAP model, due to a simultaneous spin-flip of the exciton constituents[168].

Another example of spin relaxation dynamics is CsPbI$_3$ nanocrystal studied by Strohmair *et al.* in Ref.[162] As shown in Fig. 8e-h, absorption of $\sigma+$ light excites the states with angular momentum $\Delta M_J = +1$ (from $M_{J,VB} = -1/2$ to $M_{J,CB} = +1/2$). $\sigma+$ probe monitors the population and depopulation of $\Delta M_J = +1$ states, characterized by a sharp rise and a decay to equilibrium. $\sigma-$ probe witnesses the initially unoccupied $\Delta M_J = -1$ states (from $M_{J,VB} = +1/2$ to $M_{J,CB} = -1/2$), which eventually merges with the signal of $\sigma+$ probe after ~ 10 ps. The authors found a temperature dependent spin relaxation that satisfies the Elliott-Yafet (EY) mechanism, arising from carrier-phonon interaction. In contrast to the BAP relaxation depending on electron-hole exchange interaction, the EY mechanism arises when impurities and grain boundaries scattering dominate in the spin dynamics[169], giving rise to strong temperature dependence. Giovanni *et al.* demonstrated that the spin-flip of CH$_3$NH$_3$PbI$_3$ film is governed by the EY mechanism, with a relation described by $\tau \propto T^b$, where $b$ = -0.27 for electrons and $b$ = -0.55 for holes[170]. They found relaxations of the highly J-polarized electrons and holes within 10 and 1 ps, respectively. The maximum of $\sigma+$ probe is only about 70% at 293 K and ~80% at 77 K, much lower than the total carrier population. This can be attributed to the ultrafast hole spin relaxation, which is much faster than the electron spin relaxation[120]. Spin-flip mechanisms are complicated and highly sensitive to various factors, including sample quality[170], composition[171], temperature[167], and crystal phases[172]. The realization of spintronic devices based on perovskites would greatly benefit from the investigation of the spin lifetime through helicity-resolved TAS.

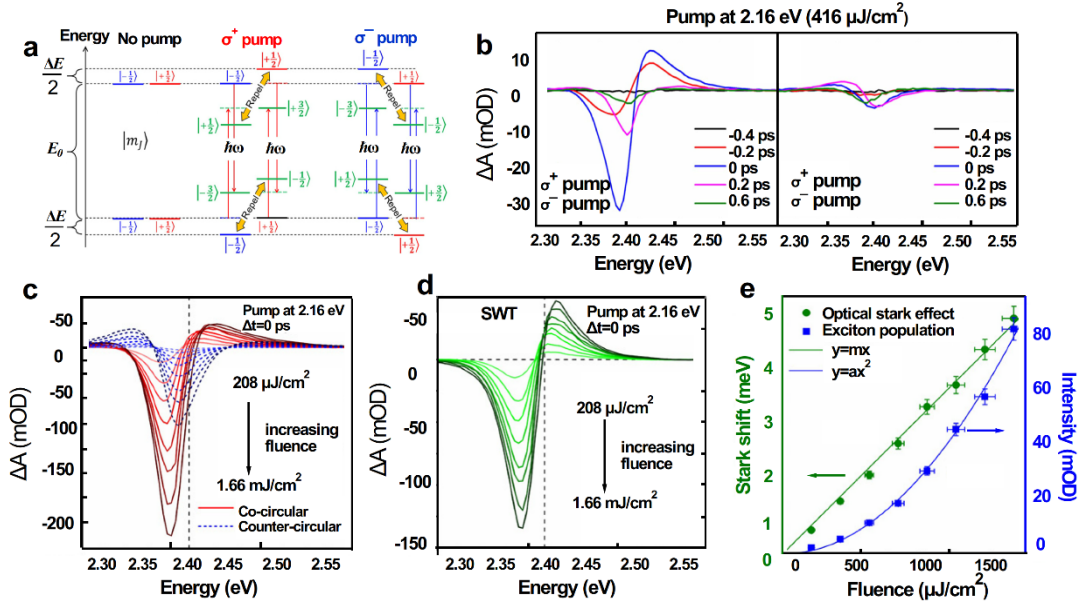

**Fig. 9** Spin-selective OSE in $(C_6H_4FC_2H_4NH_3)_2PbI_4$ perovskite thin films. (a) Schematic of the spin-selective OSE mechanism. The red (blue) arrow illustrates selective excitation with $\sigma+$ ($\sigma-$) photon. The Floquet states interact with the equilibrium with the same mJ. (b) TAS spectra of SCP ($\sigma+\sigma+$) and OCP ($\sigma+\sigma-$) pump-probe configurations at various delay time. (c) Fluence-dependent TAS spectra for co-circular (SCP, red) and counter-circular (OCP, blue) polarization pump-probe at 0 ps. (d) Resultant spectra calculated from the difference between the SCP and the OCP signals. The vertical black dashed line indicates the position of the exciton absorption peak. SWT represents spectral weight transfer. (e) Stark shift estimation as a function of pump fluence and exciton population by two-photon excitation. The Stark shift exhibits a linear relation whereas the two-photon excitation exhibits a quadratic relation with the pump fluence. Reproduced with permission from Ref. [173]. Copyright 2016, American Association for the Advancement of Science.

### 4.3 Spin-selective OSE in perovskites

As mentioned above, the OSE phenomenon arises from the hybridization of Floquet states and equilibrium excitonic states, when the circularly-polarized pump light is detuned from the resonance condition. Spin-selective OSE enables manipulation of the spin degree of freedom, facilitating its application in ultrafast spintronics[174, 175]. By employing helicity-resolved TAS with pump photon energy below exciton resonance, Giovanni *et al.* observed a strong ultrafast spin-selective OSE in $(C_6H_4FC_2H_4NH_3)_2PbI_4$ perovskite films[173]. The mechanism of spin-selective OSE is illustrated in Fig. 9a, where the Floquet states and the equilibrium occurs with the same magnetic quantum number (mJ) experience repulsion, leading to a band blue shift. Fig. 9b shows the co-circular (SCP, $\sigma+\sigma+$) and counter-circular (OCP, $\sigma+\sigma-$) pump-probe spectra. In the former case, a large photoinduced signal ($\Delta A$ arising from OSE and state filling) is observed and attributed to OSE with a spin orientation; whereas in the latter case, a minor signal presents a small state filling by two-photon excitation. Fig. 9c shows pump fluence-dependence for both SCP and OCP from 0.208 to 1.66 mJ/cm$^2$. To extract the OSE contribution and eliminate the signal from excitonic effect, the authors performed a subtraction operation by deducting the OCP signal from the SCP signal in Fig. 9c, as illustrated in Fig. 9d. Then the energy shift from OSE can be estimated from the spectral weight transfer (SWT) of the subtracted signal, $\Delta E = $ -SWT/$A(E_0)$, where $A(E_0)$ is the peak absorbance at exciton energy. Notably, the exciton bleaching peak in the OCP case exhibits fluence-dependent quadratic behavior consistent for a two-photon excitation process. In contrast, a linear relationship between the energy shift and the pump fluence can be extracted, yielding excellent agreement with the OSE theory (Fig. 9e). The authors also estimated a tuning energy of the exciton spin states by ~6.3 meV, corresponding to a Rabi energy of ~55 meV when applying a 70 T magnetic field, which is significantly larger than that in conventional systems. R. Cai *et al.* further demonstrated a biexcitonic OSE in iBA$_2$PbI$_4$ (iBA = iso-butylammonium) thin films, which allows Floquet engineering of the exciton through tuning the energy of pump photon to induce a red-shift or blue-shift[176]. It is worth mentioning that, in comparison to the aforementioned OSE in TMD MLs mostly applied at cryogenic temperatures, the spin-selective OSE in perovskites at room temperature[173] represents a crucial advancement towards the implementation of spintronic quantum devices.

# 5. Metasurfaces

In addition to the observation of chiroptical responses in TMDs and perovskites, artificially designed chiral metasurfaces also exhibit remarkable chiroptical phenomena. By employing nanofabrication techniques such as self-assembly [177], electron beam lithography [178], and glancing angle deposition [179], the spatial arrangement and geometric shapes of nanostructures can be altered to break their spatial symmetry. These chiral metasurfaces have been demonstrated to exhibit exceptional chiral properties that are unattainable in naturally occurring materials. Extensive research has been conducted for various applications such as chiral holography [180-182], Optical polarizers[16, 21, 22], and chiral light source[183, 184]. However, the lack of tunability in these chiral metasurface structures, due to fixed geometric shapes and material refractive indices at the design stage, causes limitation on their applicability in reconfigurable chiral metasurface devices. The modulation of chiroptical responses has been achieved through strategies such as adjusting geometry[185-187], refractive indices[188, 189], and experimental configurations[17]. However, challenges remain in terms of the inherent slow modulation speeds and complex fabrication processes. The modulation of the photothermal response of metasurfaces on ultrafast time scales offers a new avenue to address these challenges.

## 5.1 Control of transient chiral optical response of achiral metasurfaces

Previous studies have demonstrated that the breaking symmetry of plasmonic nanostructures can be achieved using strong femtosecond laser pulses, enabling ultrafast control over optical responses[190, 191]. Noble metal nanoresonators facilitate localized surface plasmon resonance, wherein the plasmonic dephasing process can be divided into a radiative and a non-radiative energy exchange channel. In the former case, the energy dissipates outward as photons, whereas in the latter, the energy of the incident light is absorbed after optical excitation[192, 193]. The generation of plasmonic thermal carriers occurs through non-radiative energy exchange, which can be categorized into three distinct processes: electron-electron scattering (~100 fs), electron-lattice equilibrium (~10 ps) and phonon-phonon scattering (~ns)[194-198]. Among these processes, the latter two significantly impact the permittivity of plasmonic nanostructures and offer potential for the ultrafast regulation of fully optical nanodevices. The efficiency of plasmonic thermal carrier generation has been demonstrated to be proportional to the electric field strength. Consequently, a carrier spatial distribution emerges following the spatial distribution of the plasmonic electric field, which is defined by the geometry of the nanoarray[199]. Therefore, the spatial distribution of plasmonic thermal carriers can be manipulated by designing the near-field profile and geometry of the plasmonic structures.

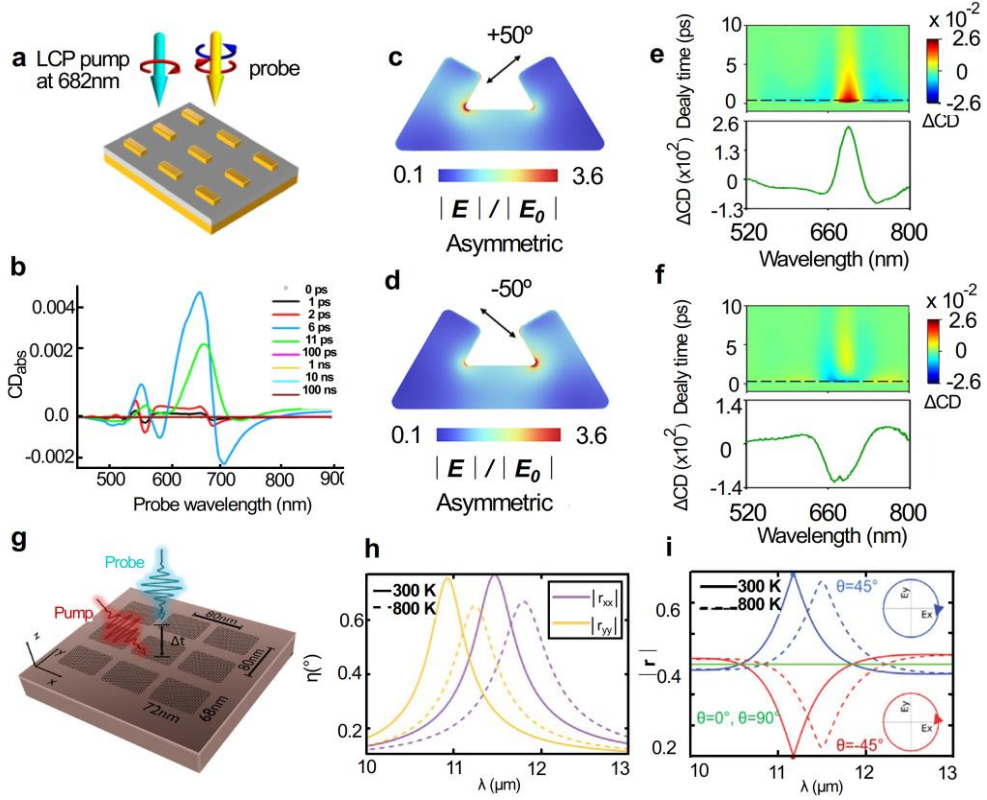

**Fig. 10** Schematic diagram of circularly polarized pump pulse light exciting an achiral metasurfaces (a) and its transient chiroptical response intensity evolution over time (b) Reproduced with permission from Ref. [200]. Copyright 2023, American Physical Society. (c) and (d) show the electric field distribution under +50º and -50º pump polarization selection, with the corresponding transient chiroptical responses shown in (e) and (f), c-f. Schematic diagram of graphene metasurface (g), its reflection amplitude as a function of electron temperature (h), and the ellipticity of the source polarization angle (e), g-i. (c-f) Reproduced with permission from Ref. [201]. Copyright 2024, National Academy of Sciences of the United States of America. (g-i) Reproduced with permission from Ref. [202]. Copyright 2024, John Wiley and Sons Inc.

To achieve ultrafast modulation of the chiroptical response of metasurfaces, an effective approach entails utilizing helicity-resolved TAS with ultrafast circularly polarized light that interacts with the metasurfaces by Avalos-Ovando *et al*., [200] as shown in Fig. 10a. By exciting the metasurface array at its resonance peak with $\sigma$- polarized light, the time-dependent evolution of the transient chiroptical response can be observed (Fig. 10b). Since this metasurface is achiral nanoarray, chiral sign can be achieved by switching to $\sigma$+ light. This work theoretically demonstrates that ultrafast modulation of chiroptical responses can be achieved in achiral metasurfaces through excitation by light helicity. Kim *et al*. further experimentally confirmed the ultrafast modulation of chiroptical responses in achiral metasurfaces using linearly polarized pulse light[201]. As shown in Fig. 10c and d, pulsed light with different polarization directions excites two diagonals of the achiral open-ring nanostructure, resulting in a large accumulation of plasmonic hot electrons at a particular corner of the structure. This transiently enhances the electric field intensity and breaks the spatial symmetry of the structure, which can be observed from the transient circular dichroism spectra, as shown in Fig. 10e and f. Obviously, it demonstrates the optical control of device chirality using linearly polarized pump light. Furthermore, Matthaiakakis *et al*. presented the possibility for achieving all-optical ultrafast dynamic control of the chiroptical response by integrating 2D materials with metasurfaces[202], as shown in Fig. 10g. This is achieved by exploiting the energy relaxation dynamics of localized surface plasmons within the metasurfaces, coupled with the instantaneous temperature rise of electrons in the two-dimensional materials (Fig. 10h and 10i). These methodologies provide novel prospects for addressing the challenges of slow modulation speed and on-site chiral sign inversion issues associated with traditional chiral metasurfaces.

## 5.2 Ultrafast polarization control of chiral metasurfaces

Ultrafast spectroscopy methods not only facilitate the observation of transient chiroptical responses in a chiral metasurface but also enable the ultrafast modulation of chiral metasurface polarization and the exploration of the physical mechanisms of chiroptical responses. For example, Tang *et al*. used helicity-resolved TAS with polarized pump-probe beams to monitor the polarization-dependent hot electron dynamics in planar chiral gold nanoclusters[203], as shown in Fig. 11a. The results reveal significant differences in transient chiroptical response intensity and hot carrier decay times, when subjected to different polarization excitations, uncovering the light control mechanisms in planar chiral gold nanoclusters on a sub-picosecond timescale (Fig. 11b). Due to the presence of inherent Ohmic losses in plasmonic nanostructures, undesirable thermal optical effects and slow thermal dissipation processes occur during pump pulse excitation, leading to poor transient chiral optical responses. To tackle this issue, silicon dielectric chiral metasurfaces for all-optical picosecond polarization switching has been proposed by Kang *et al*., [204] as shown in Fig. 11c. This combination utilizes the extremely low optical loss of silicon within the communication wavelength range, along with a high-reflectivity silver mirror backing, to achieve Fabry-Perot interference and thereby attain resonance responses with high-quality factors. By exciting carrier dynamics in the resonant cavity with pump light, dynamic control of chiroptical responses and strong ultrafast optical polarization modulation can be achieved (Fig. 11d). Moreover, despite the wider application of chiral metasurfaces across various fields, the spatiotemporal origins underlying their chiroptical responses remain unexplored. Misawa *et al*. employed time-resolved photoemission electron microscopy to investigate the near-field distribution and spatiotemporal dynamics of plasmonic modes in chiral nanorods on femtosecond scale[205], as shown in Fig. 11e. Their results indicated that the chiroptical responses of the metasurfaces are dominated by anti-symmetric (symmetric) modes of $\sigma$- ($\sigma$+) light excitation (Fig. 11f). This underscores the significant contribution of characteristic mode excitation to spatiotemporal ultrafast plasmonic chiroptical responses, therefore paving a new pathway for spatiotemporal control of chiral light-matter interactions.

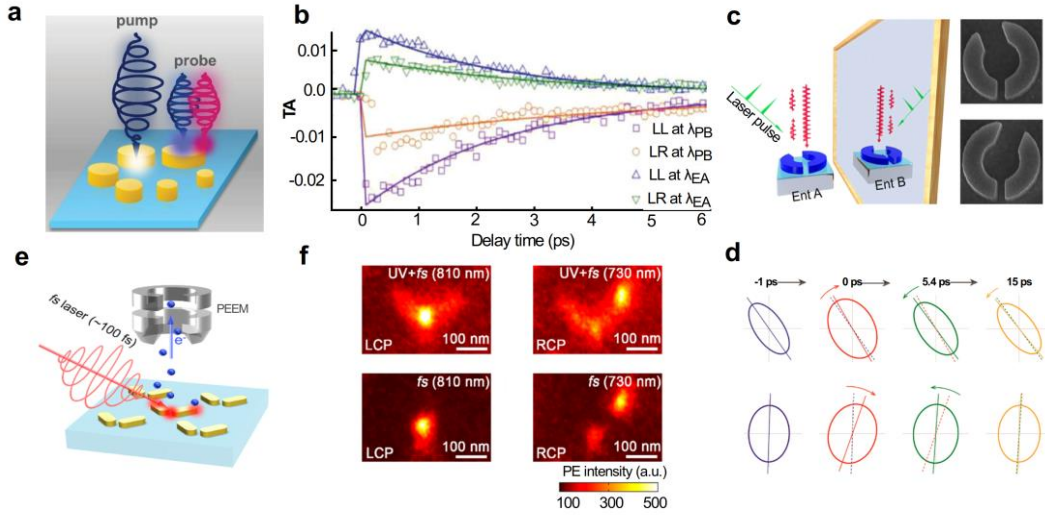

**Fig. 11** (a) Schematic diagram of ultrafast polarization measurements using a circular polarization pump beam chiral metasurface. (b) Helicity-resolved TAS of chiral metasurfaces in the excitation absorption resonance and plasmonic bleaching resonance under left circularly polarized ($\sigma$-) pump light excitation, (a-b) Reproduced with permission from Ref. [203]. Copyright 2024, John Wiley and Sons Inc. (c) Schematic diagram of all-optical switch on silicon dielectric chiral metasurface. (d) Picosecond photoswitching of light polarization at different wavelengths, c-d. Reproduced with permission from Ref. [204]. Copyright 2023, American Chemical Society. (e) Schematic diagram of time-resolved photoemission electron microscopy measurement of near-field image and spectrum of metasurface under obliquely incident light excitations. (f) Time-resolved photoemission electron microscopy images of circularly polarized light under excitation at different wavelengths, showing the dominant excitation of antisymmetric and symmetric modes under left ($\sigma$-) and right ($\sigma$+) circularly polarized excitation in the presence and absence of UV light, respectively. (e-f) Reproduced with permission from Ref. [205]. Copyright 2021, American Chemical Society.

# 6. Conclusion and outlook

In this review, we have presented the fundamental principles and experimental methodologies of helicity-resolved TAS, and summarized its applications in investigating ultrafast dynamics in low dimensional TMDs, perovskites and metasurfaces. In TMDs, valley-depolarization and valley-dependent OSE have been discovered *via* circularly polarized pump-probe configurations, arising from the spin-orbit coupling, optical selection rules, spin-valley locking, as well as intra- and inter-band scatterings. The halide perovskite materials also exhibit comparable outcomes in terms of ultrafast spin-flip and spin-selective OSE, owing to their strong spin-orbit coupling, electron-hole exchange, electron-phonon coupling, *etc*. Additionally, helicity-resolved TAS has revealed the transient optical response of metasurfaces, and explored the ultrafast control over polarization in a chiral manner.

The intriguing chiral nature of TMD MLs arises from their strong optical response at exciton resonances and valley-dependent effects. However, their responses are limited to specific photon energies due to the fixed electronic bandgaps, and the valley depolarization within a short time scale from sub-ps to few ps. The chirality response to tunable photon energies with longer depolarization time poses challenging issues that requires material development, such as TMD heterostructures. In halide perovskites, helicity-resolved TAS has unveiled several spin-quantum mechanisms originating from material composition, dimensions, and other factors. This necessitates further investigations through comprehensive experiments and theories. As for metasurfaces, existing design methods and nanofabrication processes can be further optimized to effectively enhance the transient chiral optical response intensity of plasmonic nanostructures, in order to meet the requirements for applications such as ultrafast polarization control. For example, within the existing design framework, the maximum intensity of transient chiral response can be potentially improved by using plasmonic structures with higher transient refractive index gradients.

Despite the fact that conventional TAS has emerged as a powerful tool for investigating carrier dynamics, such as energy transfer, charge transfer, and intersystem crossing relaxation, helicity-resolved TAS has been specifically developed to unravel the dynamic processes inherent chiral materials. Helicity-resolved TAS acquires data by measuring the ultrafast changes in the differential absorption between left and right circularly polarized light, facilitating the identification and characterization of short-lived intermediates with chiral characteristics formed during molecular structural transformations and chemical reactions. It enables the investigation of chirality transfer and amplification phenomena during the synthesis of chiral materials, as well as the analysis of depolarization processes of chiral optoelectronic materials upon photoexcitation. Compared to time-resolved fluorescence spectroscopy and steady-state absorption spectroscopy, one advantage of helicity-resolved TAS lies in its capability to study non-radiative recovery through many-body effects, Auger recombination, evolution of defect states, etc. Due to its exceptional sensitivity and ability to capture ultrafast dynamic processes, helicity-resolved TAS plays a crucial role in the aforementioned research fields, providing indispensable tools and methodologies for scientific research and technological developments.

There are still exciting prospects in pushing further the TAS and the related helicity-resolved TAS experimental approaches to elucidate the mechanistic understanding of the chirality nanoscale chiral materials or architectures. Due to its non-contact nature, TAS utilizes pulsed lights for exciting and tracking ultrafast dynamics, thereby facilitating convenient integration with other advanced techniques. For instance, the combination with scanning near-field microscopy (SNOM) and AFM tip-coupled light pulses, enables nanometer scale spatial resolution TAS or time-resolved SNOM, facilitating the visualization of the electric field in real space in WSe$_2$ optical waveguides[206], hyperbolic polaritons in van der Waals semiconductors[207], *etc*. Angle-resolved TAS using a combination of pump-probe techniques and an imaging spectrometer, offers a unique opportunity to investigate the ultrafast relaxation dynamics of exciton-polaritons in microcavities[208]. Revealing the pressure-related dynamics in materials, for example, pressure-induced lattice distortion and phase transition, is being facilitated by employing high-pressure TAS[209].

Acknowledgements

The authors gratefully acknowledge the strong funding support from the National Key Research and Development Program of China (grant No. 2022YFA1204700), and the National Natural Science Foundation of China (No. 92056204 and 12250710126).